\documentclass[twocolumn,aps,pra,10pt]{revtex4-2}
\usepackage{amssymb}
\usepackage{amsfonts}
\usepackage{amsmath}
\usepackage{amsxtra}
\usepackage{amscd}
\usepackage{amsthm}
\usepackage{graphicx}
\usepackage{epsf}
\usepackage{bbold}
\usepackage{xcolor}
\usepackage{array}

\newcommand{\PreserveBackslash}[1]{\let\temp=\\#1\let\\=\temp}
\newcolumntype{C}[1]{>{\PreserveBackslash\centering}p{#1}}
\newcolumntype{R}[1]{>{\PreserveBackslash\raggedleft}p{#1}}
\newcolumntype{L}[1]{>{\PreserveBackslash\raggedright}p{#1}}

\begin{document}

\title{Noninteracting tight-binding models for Fock parafermions}

\author{Edward McCann}
 \email{ed.mccann@lancaster.ac.uk}
\affiliation{Department of Physics, Lancaster University, Lancaster, LA1 4YB, United Kingdom}

\begin{abstract}
We model $p$-state Fock parafermions on a lattice in one dimension (with occupation per orbital of $0,1 , \ldots ,p-1$). For $p$ a composite number, they may be mapped to $q_m$-state parafermions where $q_m$ are the prime factors of $p$. For a Hamiltonian with a single-particle spectrum, the parafermions decompose into $q_m$-state parafermions. When $p$ is a power of two, the decomposition is into fermions. We use this to construct a parafermionic Hamiltonian for $p=4$ with a single-particle spectrum using a fermionic tight-binding model which is bilinear in creation and annihilation operators. The single-particle levels may be determined by diagonalizing a square matrix whose order scales linearly with system size, and they are the same as those of the fermionic model. We show that the intermediate statistics of the thermodynamic distribution function for the occupation numbers (known as Gentile statistics) are consistent with the mapping to fermions, and we provide an example calculation of the internal energy and heat capacity for a simple linear chain.
\end{abstract}

\maketitle

\section{Introduction}

Models of single-particle fermions are important in providing an intuitive approximate description of many physical systems as well as being the basis for perturbative calculations~\cite{fendley14,fendley24}.
Parafermions are fascinating generalizations of fermions with exclusion statistics intermediate between fermions and bosons~\cite{fendley14,cobanera14,alicea16,hutter16}, and they may provide a platform for topological quantum computing~\cite{kitaev03,nayak08,hutter16,lahtinen17}.
However, parafermions typically arise in strongly-correlated systems and generally do not have a single-particle description~\cite{fendley12,fendley14,nigro14,milsted14,jermyn14,zhuang15,sreejith16,iemini17,samajdar18,zhang19,rossini19,schmidt20,offeidanso20,lahtinen21,teixeira22b}.
There are exceptions~\cite{baxter89a,baxter89b,fendley14,alicea16,alcaraz17,alcaraz20a,alcaraz20b,alcaraz21,mastiukova22,batchelor23,mann25}, most notably Baxter's clock model~\cite{baxter89a,baxter89b,fendley14,alicea16,alcaraz17,batchelor23} which is a non-Hermitian generalization of the Ising model in one dimension (1D) with a complex, single-particle spectrum.
For $p$ state parafermions in a system with $L$ sites, there are $p^L$ many-body energies, each of the form
\begin{eqnarray}
E = \omega^{n_1} \epsilon_1 + \omega^{n_2} \epsilon_2 + \ldots + \omega^{n_L} \epsilon_L ,
\end{eqnarray}
where $\omega = \exp (2 \pi i / p)$, $(n_1, n_2, \ldots , n_L)$ are parafermion occupation numbers $n_j = 0,1,\ldots , p-1$, and $\epsilon_j$ are real single-particle energy levels found by diagonalizing a square matrix whose order scales linearly with the system size $L$. Moreover, the levels $\epsilon_j$ for arbitrary $p$ are related to those of the Ising model (with $p=2$)~\cite{baxter89a,baxter89b,fendley14,alicea16,alcaraz17}.

In this paper, we consider noninteracting tight-binding models for four-state ($p=4$) Fock parafermions~\cite{cobanera14} in 1D with a real single-particle spectrum. Four-state Fock parafermions~\cite{cobanera14} are indistinguishable quantum particles with an associated Fock space, and creation and annihilation operators where each orbital~\cite{orbitalcomment} may have an occupancy of $0$, $1$, $2$ or $3$. For a system with $L$ orbitals, there are $4^L$ many-body energies, each of the form
\begin{eqnarray}
E = n_1 \epsilon_1 + n_2 \epsilon_2 + \ldots + n_L \epsilon_L , \label{pspectrum}
\end{eqnarray}
where $(n_1, n_2, \ldots , n_L)$ are parafermion occupation numbers $n_j = 0,1,2,3$, and $\epsilon_j$ are real single-particle energy levels found by diagonalizing a square matrix whose order scales linearly with the system size $L$.
At finite temperature, the mean occupation number of a single-particle level $\langle n_{\ell} \rangle$ for $p$-state parafermions with spectrum~(\ref{pspectrum}) is an example of statistics intermediate between the Fermi-Dirac and Bose-Einstein distributions, also known as Gentile statistics~\cite{gentile40,gentile42}.

Single-particle tight-binding models of fermions are solvable because the number operator and the Hamiltonian are bilinear functions of creation and annihilation operators~\cite{asboth16,mccann23,fendley24}. For Fock parafermions~\cite{cobanera14}, however, there is a nonlinear relation between the number operator and the creation and annihilation operators, as well as nonlinear onsite commutation relations. Hence, it is not possible to solve an arbitrary tight-binding model for Fock parafermions~\cite{cobanera14,xu17,calzona18,rossini19,mahyaeh20,camacho22,bahovadinov22}, even if the Hamiltonian is bilinear, e.g., with single-particle hopping only.

However, there are mappings from spin-$1/2$ fermions to four-state ($p=4$) parafermions~\cite{kohmoto81,yamanaka94,hutter15,yu15,meichanetzidis18,calzona18,chew18,bomantara21,teixeira22,traverso23,osvath24,safwan25}.
Here, we modify a previously considered mapping~\cite{calzona18} so that the parafermion number operator $\hat{N}_j$ can be written as a linear combination $\hat{N}_j = \hat{n}_{j \uparrow} + 2 \hat{n}_{j \downarrow}$ of number operators for spin up $\hat{n}_{j \uparrow}$ and spin down $\hat{n}_{j \downarrow}$ fermions. This allows the occupations of $0$, $1$, $2$ or $3$ to be described by fermions with occupations of $0$ or $1$.
Thus, energies in a single-particle spectrum~(\ref{pspectrum}) for four-state parafermions may be written as a linear combination of single-particle spectra for fermions, namely
\begin{eqnarray}
E &=& n_{1 \uparrow} \epsilon_1 + n_{2 \uparrow} \epsilon_2 + \ldots + n_{L \uparrow} \epsilon_L \nonumber \\
&& \qquad + 2 (n_{1 \downarrow} \epsilon_1 + n_{2 \downarrow} \epsilon_2 + \ldots + n_{L \downarrow} \epsilon_L) . \label{esplit}
\end{eqnarray}
Four-state parafermions with such a spectrum can be considered to decompose into two species of fermions.

With this realization, it is possible to construct Hamiltonians for four-state parafermions with the single-particle spectrum~(\ref{pspectrum}) using any single-particle tight-binding model for fermions which is bilinear in fermionic creation and annihilation operators. Here, we describe an example in 1D with arbitrary onsite energies and nearest-neighbor hopping parameters. Such a Hamiltonian conserves the number of spin up and of spin down fermions as well as the number of parafermions.
We show that thermodynamic properties of the noninteracting four-state parafermions, particularly the intermediate statistics of the occupation distribution function (also known as Gentile statistics~\cite{gentile40,gentile42}), are consistent with the mapping to spin-$1/2$ fermions.
Finally, we describe the generalization to $p$-state parafermions~\cite{bondesan13,cobanera14,moran17}: When $p$ is a composite number, occupation numbers may be written as a linear combination of occupation numbers for $q_m$-state parafermions, where $q_m$ are the prime factors of $p$. Particularly, when $p$ is a power of two, parafermions with a single-particle spectrum decompose into fermions.

\section{Tight-binding model}

We consider a tight-binding model consisting of $L$ sites, arbitrary onsite energies $u_j$ for $j=1,2,\ldots , L$, and nearest-neighbor hopping  characterized by arbitrary parameters $t_j$ for $j=1,2,\ldots , L-1$. For simplicity, we consider real tight-binding parameters, giving an Hermitian Hamiltonian for four-state ($p=4$) parafermions with open boundary conditions,
\begin{eqnarray}
H &=& \sum_{j=1}^L u_j \hat{N}_j + \sum_{j=1}^{L-1} t_j \big[ c_j^{\dagger} \hat{\Gamma}_j \hat{\Gamma}_{j+1}  c_{j+1} (-i)^{\hat{N}_{j}}  \nonumber \\
&& \qquad \qquad \qquad + 2 c_j^{\dagger 2} c_{j+1}^2 (-1)^{\hat{N}_{j}} + \mathrm{H.c.} \big] , \label{ham1}
\end{eqnarray}
where $\hat{\Gamma}_j = 1 - c_j^{\dagger} c_j + c_j^{\dagger 2} c_j^2$.
Parafermion creation $c_j^{\dagger}$ and annihilation $c_j$ operators act on Fock states~\cite{cobanera14} as
\begin{eqnarray*}
c_j^{\dagger} | n_1 , \ldots , n_j , \ldots , n_L \rangle \!&=&\!
(-i)^{\sum_{\ell < j} n_{\ell}}  | n_1 , \ldots , n_j + 1 , \ldots , n_L \rangle , \\
c_j | n_1 , \ldots , n_j , \ldots , n_L \rangle \!&=&\!
i^{\sum_{\ell < j} n_{\ell}}  | n_1 , \ldots , n_j - 1 , \ldots , n_L \rangle ,
\end{eqnarray*}
where $n_j = 0,1,2,3$ is an eigenvalue of the number operator $\hat{N}_j$~\cite{cobanera14},
\begin{eqnarray}
\hat{N}_j | n_1 , \ldots , n_j , \ldots , n_L  \rangle =
n_j | n_1 , \ldots , n_j , \ldots , n_L  \rangle , \label{pn1}
\end{eqnarray}
which is given by
\begin{eqnarray}
\hat{N}_j = c_j^{\dagger} c_j + c_j^{\dagger 2} c_j^2 + c_j^{\dagger 3} c_j^3 . \label{pn2}
\end{eqnarray}
Operators on different sites commute up to a phase~\cite{cobanera14},
\begin{eqnarray*}
c_{\ell}^{\dagger} c_j^{\dagger} = i c_j^{\dagger} c_{\ell}^{\dagger} ; \quad
c_{\ell} c_j = i c_j c_{\ell} ; \quad
c_{\ell} c_j^{\dagger} = - i c_j^{\dagger} c_{\ell} \quad \mathrm{for}\,\, \ell < j .
\end{eqnarray*}
whereas, on the same site,
\begin{eqnarray*}
c_j^{\dagger m} c_j^m+ c_j^{4-m} c_j^{\dagger (4-m)} = 1  \quad \mathrm{for}\,\, m = 1 , \ldots , 3 .
\end{eqnarray*}
In the Hamiltonian~(\ref{ham1}), the combination of terms, including string phase factors such as $(-i)^{\hat{N}_{j}}$, ensures that it is local in space~\cite{cobanera14,calzona18}, i.e., $H$ could be written in terms of parafermionic versions of hard-core bosons~\cite{cobanera14} without any string factors.
The operator $\hat{\Gamma}_j$ restricts single-particle hopping so that it is not possible to perform a single hop off a site that is doubly occupied or for a single hop to create a doubly occupied site. However, these states can generally be accessed by coherent hopping of two particles [the second hopping term in Eq.~(\ref{ham1})].

The Hamiltonian~(\ref{ham1}) may be solved by mapping~\cite{supplementary} the parafermions to itinerant spinful fermions, $\sigma = \{ \uparrow , \downarrow \}$, with creation and annihilation operators $f_{j \sigma}^{\dagger}$ and $f_{j \sigma}$,
\begin{eqnarray}
f_{j \uparrow} &=& i^{\sum_{\ell < j} \hat{N}_{\ell}} \hat{\Gamma}_j  c_{j} , \label{budef} \\
f_{j \downarrow} &=& (-1)^{\sum_{\ell \geq j} \hat{n}_{\ell \uparrow}} (-1)^{\sum_{\ell < j} \hat{n}_{\ell \downarrow}} c_{j}^2 , \label{bddef} \\
c_j &=& (-i)^{\sum_{\ell < j} \hat{N}_{\ell}} f_{j \uparrow} \nonumber \\
&& \quad + i^{\sum_{\ell < j} \hat{n}_{\ell \uparrow}} (-1)^{\sum_{\ell > j} \hat{n}_{\ell \uparrow}} f_{j \uparrow}^{\dagger} f_{j \downarrow} , \label{cbdef}
\end{eqnarray}
where $\hat{n}_{j \sigma}$ is the number operator for fermions on site $j$,
\begin{eqnarray}
\hat{n}_{j \uparrow} &=& f_{j \uparrow}^{\dagger} f_{j \uparrow}
= c_j^{\dagger} c_j - c_j^{\dagger 2} c_j^2 + c_j^{\dagger 3} c_j^3 , \label{nup} \\
\hat{n}_{j \downarrow} &=& f_{j \downarrow}^{\dagger} f_{j \downarrow}
= c_j^{\dagger 2} c_j^2 . \label{ndown}
\end{eqnarray}

The annihilation operators $f_{j \sigma}$ for spinful fermions act on Fock states with $n_{j \sigma} = 1$ as
\begin{eqnarray*}
&& f_{j \uparrow} | \ldots , n_{j \uparrow} , \ldots \rangle =
(-1)^{\sum_{\ell < j} \hat{n}_{\ell \uparrow}}  | \ldots , n_{j \uparrow}-1 , \ldots \rangle , \\
&& f_{j \downarrow} | \ldots , n_{j \downarrow} , \ldots \rangle = \nonumber \\
&& \qquad (-1)^{\sum_{\ell =1}^L \hat{n}_{\ell \uparrow}} (-1)^{\sum_{\ell < j} \hat{n}_{\ell \downarrow}} | \ldots , n_{j \downarrow}-1 , \ldots \rangle ,
\end{eqnarray*}
where $n_{j \sigma}$ is an eigenvalue of the number operator $\hat{n}_{j \sigma}$.
These operators obey the usual fermionic anticommutation relations including anticommutation of different spins, $\{ f_{j \sigma} , f_{\ell \sigma^{\prime}}^{\dagger} \} = \delta_{j \ell} \delta_{\sigma \sigma^{\prime}}$.
The mapping relates the onsite parafermion basis $\{ |0\rangle , |1\rangle , |2\rangle , |3\rangle \}$ to the action of spinful fermions on the vacuum,
\begin{eqnarray*}
|0\rangle \equiv |0\rangle ; \quad
|1\rangle \equiv f_{j \uparrow}^{\dagger} |0\rangle ; \quad
|2\rangle \equiv f_{j \downarrow}^{\dagger} |0\rangle ; \quad
|3\rangle \equiv f_{j \uparrow}^{\dagger} f_{j \downarrow}^{\dagger} |0\rangle ,
\end{eqnarray*}
which is a modification of the mapping described in~\cite{calzona18}.

With the fermionic number operators~(\ref{nup}) and (\ref{ndown}), the parafermionic number operator~(\ref{pn2}) may be expressed as
\begin{eqnarray}
\hat{N}_j = f_{j \uparrow}^{\dagger} f_{j \uparrow} + 2 f_{j \downarrow}^{\dagger} f_{j \downarrow} . \label{nfs}
\end{eqnarray}
This mapping allows the Hamiltonian to be written as a sum of quadratic terms,
\begin{eqnarray}
H = \Phi_{\uparrow}^{\dagger} {\cal H} \Phi_{\uparrow}
+ 2 \Phi_{\downarrow}^{\dagger} {\cal H} \Phi_{\downarrow} , \label{ham2}
\end{eqnarray}
where $\Phi_{\sigma}^{\dagger} = \begin{pmatrix}
f_{1 \sigma}^{\dagger} & f_{2 \sigma}^{\dagger} & \ldots & f_{L \sigma}^{\dagger} \end{pmatrix}$
and ${\cal H}$ is a $L \times L$ matrix,
\begin{eqnarray}
{\cal H} = \begin{pmatrix}
u_1 & t_1 & 0 & 0 & \hdots \\
t_1 & u_2 & t_2 & 0 & \hdots \\
0 & t_2 & u_3 & t_3 & \hdots \\
0 & 0 & t_3 & u_4 & \hdots \\
\vdots & \vdots & \vdots & \vdots & \ddots
\end{pmatrix} . \label{hgenmat}
\end{eqnarray}
The linear relationship~(\ref{nfs}) between the parafermion number operator and the fermion number operators is the reason why the model is solvable with a single-particle spectrum. On diagonalizing the matrix ${\cal H}$, linear combinations of $f_{j \sigma}^{\dagger}$ and $f_{j \sigma}$ create number operators for fermion eigenstates which are combined, by the form of the Hamiltonian~(\ref{ham2}), into number operators for parafermion eigenstates.
Thus, the many-body energy spectrum of Hamiltonian~(\ref{ham1}) consists of $4^L$ energies, each of which may be expressed as in Eq.~(\ref{pspectrum})
where $\epsilon_{\ell}$ are single-particle energy levels with $\ell = 1,2,\ldots, L$, i.e., eigenvalues of the matrix ${\cal H}$, and $n_{\ell} = 0,1,2,3$ are parafermion occupation numbers~\cite{supplementary}.

For simplicity, we have considered a model with nearest-neighbor hopping only. However, the construction of parafermion models with a single-particle spectrum~(\ref{pspectrum}) may be generalized to any fermion model which is bilinear in fermion creation and annihilation operators.
An example is given in the Supplemental Material~\cite{supplementary} with next-nearest-neighbor hopping, giving a non-local parafermion Hamiltonian.
A second example in the Supplemental Material~\cite{supplementary} generalizes the approach to mean-field models of a superconductor that may be solved using a Bogoliubov transformation. Specifically, we construct a parafermion counterpart of the Kitaev superconducting chain~\cite{kitaev01}, and we show that the ground state in the topological phase is fourfold degenerate, with each ground state distinguished by the fourfold parafermion occupation numbers of Majorana edge modes.

\section{Translational invariance}

When the parameters in a Hamiltonian have translational invariance, Fourier transformation of parafermion operators is generally not useful because of complicated commutation relations in $k$ space~\cite{fendley12,li15}.
However, with the mapping~(\ref{nfs}) and~(\ref{ham2}), we can write parafermion eigenvalues and eigenstates in terms of known fermion ones. 
As an example, consider a simple linear chain in 1D with one orbital per unit cell, $u_j = u$ and $t_j = t$ for all $j$ in the Hamiltonian~(\ref{ham1}) and the matrix~(\ref{hgenmat}). The eigenvalues for a fermion system with open boundary conditions are $\epsilon_k = u + 2 t \cos (ka)$ where $ka = m \pi / (L+1)$ with index $m = 1,2,\ldots , L$. The matrix has eigenstates $\psi_k (j) = \sqrt{2/(L+1)} \sin (kaj)$, and the fermionic Hamiltonians may be diagonalized as $H_{\sigma} = \sum_k \epsilon (k) \tilde{f}_{k \sigma}^{\dagger} \tilde{f}_{k \sigma}$ where $\tilde{f}_{k \sigma} = \sum_j \psi_k (j) f_{j \sigma}$.
With $H = H_{\uparrow} + 2 H_{\downarrow}$, then the parafermion Hamiltonian may be written as $H = \sum_k \epsilon (k) \hat{N}_k$ where $\hat{N}_k = \hat{n}_{k \uparrow } + 2\hat{n}_{k \downarrow }$ for $\hat{n}_{k \sigma } = \tilde{f}_{k \sigma}^{\dagger} \tilde{f}_{k \sigma}$.

\section{Parafermion distribution function}
Given a single-particle energy spectrum~(\ref{pspectrum}), one would like to know the occupation of levels as a function of temperature $T$ and chemical potential $\mu$.
The mean occupation number of a single-particle level $\langle n_{\ell} \rangle$ for four-state parafermions with spectrum~(\ref{pspectrum}) is an example of statistics intermediate between the Fermi-Dirac and Bose-Einstein distributions, also known as Gentile statistics~\cite{gentile40,gentile42}.
The grand canonical partition function $Z$ may be represented as $Z = \prod_{\ell} Z_{\ell}$ where the partition function for a given single-particle level is $Z_{\ell} = \sum_{n_{\ell}} \exp [\beta (\mu - \epsilon_{\ell}) n_{\ell}]$, and $\beta = 1 / k_{\mathrm{B}} T$ is inverse temperature~\cite{statphys,dutt94,stoilova20}.
For four-state parafermions,
\begin{eqnarray}
Z_{\ell} = 1 + x^{-1} + x^{-2} + x^{-3} ; \qquad
x = e^{\beta (\epsilon_{\ell} - \mu)} .
\end{eqnarray}
The mean occupation number is given~\cite{gentile40,gentile42} by
\begin{eqnarray}
\langle n_{\ell} \rangle = \frac{1}{\beta} \frac{\partial \ln Z_{\ell}}{\partial \mu} =
\frac{3 + 2x + x^2}{(1+x)(1+x^2)} .
\end{eqnarray}
This may be expressed as a sum of two fermionic terms,
\begin{eqnarray}
\langle n_{\ell} \rangle = \langle n_{\ell \uparrow} \rangle + 2\langle n_{\ell \downarrow} \rangle , \label{thermosum}
\end{eqnarray}
in agreement with the expectations of the mapping from parafermions to spin-1/2 fermions, Eq.~(\ref{nfs}).
Here,
$\langle n_{\ell \sigma} \rangle = 1/[e^{\beta_{\sigma} (\epsilon_{\ell} - \mu )} + 1]$
where $\beta_{\uparrow} = \beta$ and $\beta_{\downarrow} = 2\beta$, i.e., the down spin component $\langle n_{\ell \downarrow} \rangle$ has a Fermi-Dirac distribution, but at a lower effective temperature of $T/2$.
At low temperature, $\langle n_{\ell} \rangle \approx 3$ for $\epsilon_{\ell} \ll \mu$, and $\langle n_{\ell} \rangle \approx 0$ for $\epsilon_{\ell} \gg \mu$~\cite{gentile40,gentile42,stoilova20}.
Owing to the lower effective temperature of the down spin component, $\langle n_{\ell} \rangle \approx \exp [ - \beta (\epsilon_{\ell} - \mu) ]$ for $\beta (\epsilon_{\ell} - \mu) \gg 1$, recovering Maxwell-Boltzmann statistics.

The variance $\langle (\Delta n_{\ell})^2 \rangle$ in the occupation number of a single-particle level may be determined with
$\langle (\Delta n_{\ell})^2 \rangle = (1/\beta) \partial \langle n_{\ell} \rangle / \partial \mu$~\cite{statphys},
\begin{eqnarray}
\langle (\Delta n_{\ell})^2 \rangle = \langle n_{\ell \uparrow} \rangle (1 - \langle n_{\ell \uparrow} \rangle) + 4 \langle n_{\ell \downarrow} \rangle (1 - \langle n_{\ell \downarrow} \rangle) . \label{var}
\end{eqnarray}
As with the Fermi-Dirac distribution, $\langle (\Delta n_{\ell})^2 \rangle \approx 0$ for $\epsilon_{\ell} \ll \mu$ and for $\epsilon_{\ell} \gg \mu$ at low temperature, and it has its maximal value at $\epsilon_{\ell} = \mu$ where $\langle (\Delta n_{\ell})^2 \rangle_{\mathrm{max}} = 5/4$. This is larger than that of the Fermi-Dirac distribution, $\langle (\Delta n_{\ell})^2 \rangle_{\mathrm{max}} = 1/4$, as expected for statistics intermediate between those of fermions and bosons.

The mean parafermionic occupation number~(\ref{thermosum}) determines thermodynamic properties of parafermionic systems with single-particle spectra~(\ref{pspectrum}) such as the internal energy per unit length $u_E = \int_{-\infty}^{\infty} \langle n (\epsilon ) \rangle g (\epsilon) \epsilon d \epsilon$ and the specific heat capacity $c_V = \partial u_E / \partial T$~\cite{statphys}.
Here $\langle n (\epsilon ) \rangle$ is the parafermionic occupation number~(\ref{thermosum}) in the limit of continuous energy $\epsilon$, and $g (\epsilon)$ is the energy density of states per unit length.
With the parafermion occupation number~(\ref{thermosum}),
$u_E (T) = u_0 (T) + 2 u_0 (T/2)$, where $u_0 (T)$ is the internal energy of a system of nondegenerate fermions.
Specifically,
\begin{eqnarray}
u_E &=& \int_{-\infty}^{\infty} \frac{g (\epsilon) \epsilon d \epsilon}{e^{\beta (\epsilon_{\ell} - \mu)} + 1 } + 2 \int_{-\infty}^{\infty} \frac{g (\epsilon) \epsilon d \epsilon}{e^{2\beta (\epsilon_{\ell} - \mu)} + 1 } . \label{pu}
\end{eqnarray}
For $T=0$, $u_E = 3 \int_{-\infty}^{\epsilon_{\mathrm{F}}} g (\epsilon) \epsilon d \epsilon$, where $\epsilon_{\mathrm{F}} = \mu (T=0)$, which is three times that of nondegenerate fermions~\cite{statphys}.
The specific heat of parafermions is $c_V (T) = c_0 (T) + c_0 (T/2)$, where $c_0 (T)$ is the specific heat of nondegenerate fermions.
At low temperature, the linear-in-temperature part of the specific heat 
$c_{V} \approx (\pi^2 / 2) k_{\mathrm{B}}^2 T g ( \epsilon_{\mathrm{F}} )$ is $3/2$ that of nondegenerate fermions~\cite{statphys}.

\begin{figure}[t]
\includegraphics[scale=0.32]{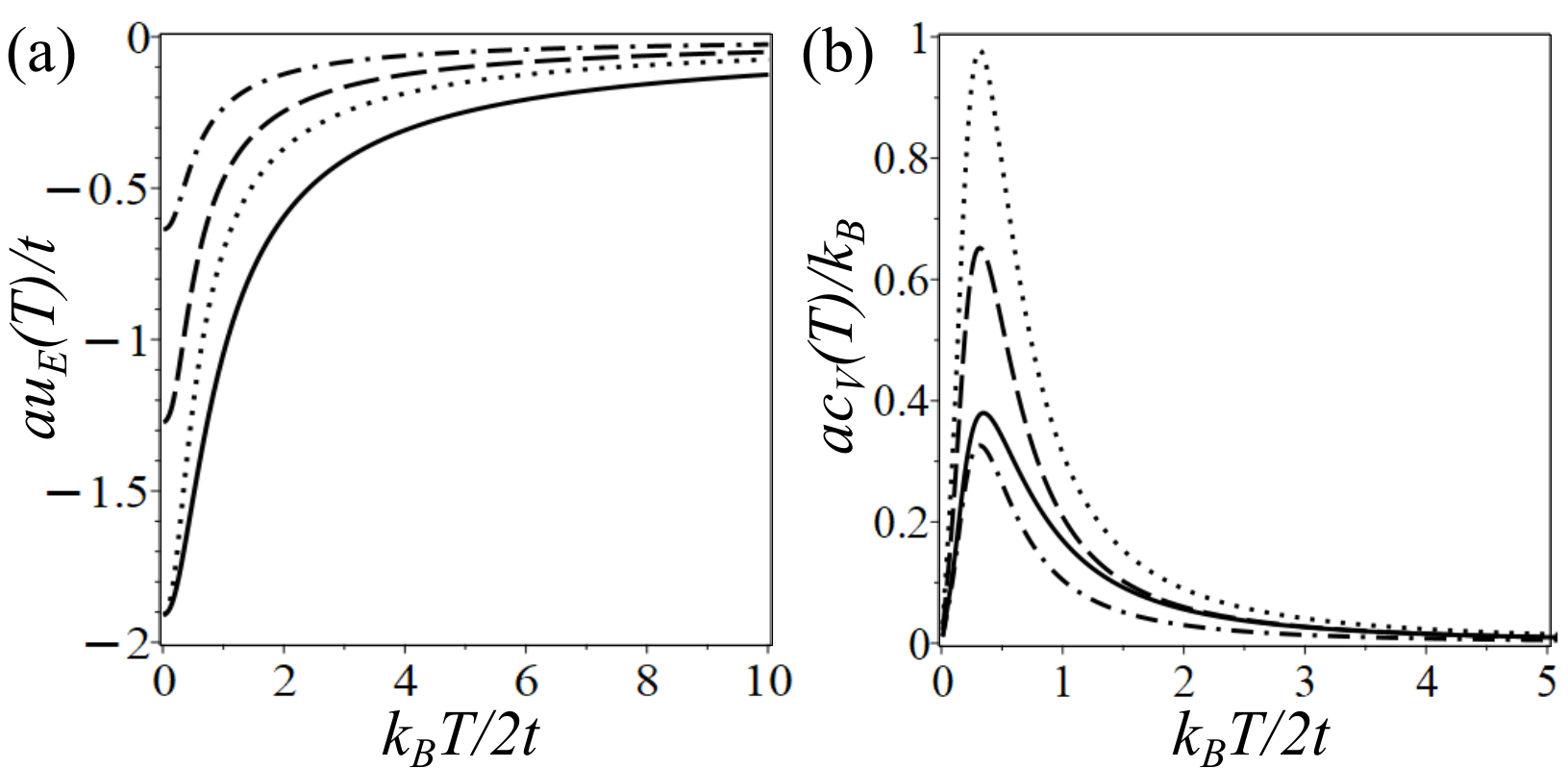}
\caption{(a) Internal energy per unit length $u_E$ and (b) specific heat capacity $c_V$ plotted as a function of temperature $T$, for a linear chain with nearest-neighbor hopping $t$ with dispersion $\epsilon (k) = 2t \cos (ka)$ where $a$ is the lattice constant. Curves show the contribution of four-state parafermions (solid), nondegenerate fermions (dash-dotted), twofold degenerate fermions (dashed), and threefold degenerate fermions (dotted).
For all plots, the chemical potential $\mu = 0$. Data is obtained by numerical evaluation of $u_E = \int_{-\infty}^{\infty} \langle n (\epsilon ) \rangle g (\epsilon) \epsilon d \epsilon$ and $c_V = \partial u_E / \partial T$.
}\label{figtherm}
\end{figure}

As a specific example, consider the simple linear chain with one orbital per unit cell, zero onsite energy ($u=0$), nearest-neighbor hopping $t$, and a single band $\epsilon_k = 2t \cos (ka)$ for $L \gg 1$. The density of states is $g (\epsilon ) = 1 / ( \pi a \sqrt{4t^2 - \epsilon^2} )$ for $|\epsilon | < 2t$.
Figure~\ref{figtherm}(a) shows $u_E (T)$ and Fig.~\ref{figtherm}(b) shows $c_V (T)$ when the chemical potential is at the center of the band, $\mu = 0$, determined by numerical evaluation of Eq.~(\ref{pu}) and $c_V = \partial u_E / \partial T$. The properties of parafermions (solid lines) are compared to those of fermions with degeneracies ranging from one to three. 
The internal energy of parafermions, Fig.~\ref{figtherm}(a), coincides with that of threefold-degenerate fermions at $T=0$, but the temperature dependence is different according to $u_E (T) = u_0 (T) + 2 u_0 (T/2)$ with the magnitude of the parafermion energy being larger for $T>0$.
The heat capacity of parafermions, Fig.~\ref{figtherm}(b), is also distinct from that of fermions for $T>0$ according to $c_V (T) = c_0 (T) + c_0 (T/2)$. All the curves show a linear-in-$T$ dependence at low $T$, but with different gradients.

The parafermion Hamiltonian~(\ref{ham1}) conserves the total number of spin up fermions and of spin down fermions, and, as a consequence, it is block diagonalized in the many-body basis of atomic orbitals $ | n_1 , n_2, \ldots , n_L \rangle$. Each energy $E$ in the single-particle parafermion energy spectrum~(\ref{pspectrum}) may be written as a sum of two parts corresponding to two separate fermionic species~(\ref{esplit}).
Here, we show that interpreting the spectra as consisting of two separate fermionic species is consistent with the thermodynamics of single-particle parafermions, specifically we consider the internal energy~(\ref{pu}).
The mean occupation numbers $\langle n_{\ell \uparrow} \rangle$ and $\langle n_{\ell \downarrow} \rangle$ follow the Fermi-Dirac distribution with chemical potentials $\mu_{\uparrow}$ and $\mu_{\downarrow}$, respectively.
We relate the chemical potentials at zero temperature to the number of particles, assuming that $N \approx 3 N_{\uparrow} \approx 3 N_{\downarrow}$ in the thermodynamic limit,
where $N$ is the total number of parafermions and $N_{\uparrow}$ ($N_{\downarrow}$) is the total number of spin up (down) fermions.
For the spin up fermions, $g_{\uparrow} (\epsilon ) = g( \epsilon)$ where $g( \epsilon)$ is the parafermion density of states [as they have the same spectrum according to Eq.~(\ref{ham2})], and $N_{\uparrow} = L \int_{-\infty}^{\mu_{\uparrow}} g (\epsilon ) d \epsilon$.
For spin down, $g_{\downarrow} (\epsilon ) = (1/2) g ( \epsilon / 2)$ and $N_{\downarrow} = L \int_{-\infty}^{\mu_{\downarrow}/2} g ( \epsilon) d \epsilon$.
Finally, for the parafermions with chemical potential $\mu$,
$N = 3L \int_{-\infty}^{\mu} g (\epsilon ) d \epsilon$.
Comparing these expressions with $N \approx 3 N_{\uparrow} \approx 3 N_{\downarrow}$ gives
$\mu_{\uparrow} \approx \mu_{\downarrow} / 2 \approx \mu$.
Thus, the internal energy is given by
\begin{eqnarray}
u_E &=& \int_{-\infty}^{\infty} \langle n_{\uparrow} (\epsilon) \rangle g_{\uparrow} (\epsilon) \epsilon d \epsilon
+ \int_{-\infty}^{\infty} \langle n_{\downarrow} (\epsilon^{\prime}) \rangle g_{\downarrow} (\epsilon^{\prime}) \epsilon^{\prime} d \epsilon^{\prime} , \nonumber \\
&=& \int_{-\infty}^{\infty} \frac{g (\epsilon) \epsilon d \epsilon}{e^{\beta (\epsilon - \mu)} + 1 }
+ \frac{1}{2} \int_{-\infty}^{\infty} \frac{g (\epsilon^{\prime} / 2) \epsilon^{\prime} d \epsilon^{\prime}}{e^{\beta (\epsilon^{\prime} - 2\mu)} + 1 } .
\end{eqnarray}
Substituting $\epsilon^{\prime} = 2 \epsilon$ recovers the parafermion result~(\ref{pu}).

\section{Generalization to $p$-state parafermions}

When $p$ is a composite number, we consider the factorization of $p$ into prime factors $q_m$ (including repeated factors) as
\begin{eqnarray}
p = q_1 q_2 \cdots q_k ,
\end{eqnarray}
where the prime factors are placed in ascending order, $q_m \leq q_{\ell}$ for $m< \ell$. By counting the possible occupations $0,1, \ldots , p-1$, an occupation number for $p$-state parafermions $n^{(p)}$ may be written as a linear combination of occupation numbers for $q_m$-state parafermions,
\begin{eqnarray}
n^{(p)} = n_1^{(q_1)} + \sum_{m=2}^k \left( \prod_{\ell =1}^{m-1} q_{\ell} \right) n_m^{(q_m)} ,
\end{eqnarray}
where the subscript is used to distinguish different species (which is needed for repeated factors), and $k$ is the number of prime factors including repeated factors.
We expect there is a corresponding mapping of creation and annihilation operators and of number operators, and such mappings for $p=6$ and $p=8$ are given in the Supplemental Material~\cite{supplementary}.
For $p$ composite and a Hamiltonian with a single-particle spectrum~(\ref{pspectrum}), the parafermions decompose into $q_m$-state parafermions.
Hence, a $p$-state parafermion satisfying Gentile statistics~\cite{gentile40,gentile42} is a composite particle when $p$ is a composite number.
For the special case when $p$ is a power of two, the parafermions decompose into fermions, and it is possible to construct parafermionic Hamiltonians with a single-particle spectrum, as shown here for $p=4$ using Eq.~(\ref{ham2}).
Note that one could consider any solvable fermionic model, i.e., not necessarily of the form Eq.~(\ref{ham2}), and use the mapping to write it in a four-state parafermion representation with a known energy spectrum. However, it generally wouldn't have the single-particle parafermionic spectrum~(\ref{pspectrum}).

\section{Conclusions}

For $p$-state Fock parafermions, when $p$ is a power of two, and for a Hamiltonian with a single-particle spectrum, the parafermions decompose into fermions. We use this to construct a parafermionic Hamiltonian for $p=4$ with a single-particle spectrum using a fermionic tight-binding model which is bilinear in creation and annihilation operators.
The thermodynamic distribution function for noninteracting four-state parafermions~(\ref{thermosum}), known as Gentile statistics~\cite{gentile40,gentile42}, is consistent with the mapping, but yields a down spin component with a Fermi-Dirac distribution at a lower effective temperature $T/2$.

The mapping and decomposition may be extended to $p$-state Fock parafermions when $p$ is a composite number, and mappings for $p=6$ and $p=8$ are given in the Supplemental Material~\cite{supplementary}.
We speculate that the mapping may provide an avenue for experimental simulation of parafermionic systems, e.g., using separate fermionic systems. Further theoretical studies should consider the properties of weakly-interacting parafermions, for which the decomposition would break down. Another open question is whether it is possible to realize models with the single-particle spectrum~(\ref{pspectrum}), and, hence, satisfying Gentile statistics~\cite{gentile40,gentile42}, when $p$ is prime.

\begin{acknowledgments}
The author thanks J. Barnett, F. Schindler, and H. Schomerus for helpful discussions.
\end{acknowledgments}

\section*{Data availability}
The data that support the findings of this article are openly available~\cite{datanote}.

\onecolumngrid
\clearpage
\begin{center}
\textbf{\large Supplementary material: Noninteracting tight-binding models for Fock parafermions}
\end{center}

\setcounter{section}{0}
\setcounter{equation}{0}
\setcounter{figure}{0}
\setcounter{table}{0}
\setcounter{page}{1}
\makeatletter
\renewcommand{\theequation}{S\arabic{equation}}
\renewcommand{\thefigure}{S\arabic{figure}}
\renewcommand{\bibnumfmt}[1]{[S#1]}
\renewcommand{\citenumfont}[1]{S#1}

\begin{itemize}
\item[I] The mapping from $p=4$ parafermions to spinful fermions
\item[II] Exact diagonalization of the tight-binding model
\item[III] Model with next-nearest-neighbor hopping
\item[IV] Kitaev superconducting chain
\item[V] Mapping for $p=6$
\item[VI] Mapping for $p=8$
\end{itemize}

\section{The mapping from $p=4$ parafermions to spinful fermions}

For the mapping given in the main text, Eqs.~(7), (8), (9), the annihilation operators $f_{j \sigma}$ for spinful fermions act on Fock states with $n_{j \sigma} = 1$ as
\begin{eqnarray}
f_{j \uparrow} | \ldots , n_{j \uparrow} , \ldots \rangle &=&
(-1)^{\sum_{\ell < j} \hat{n}_{\ell \uparrow}}  | \ldots , n_{j \uparrow}-1 , \ldots \rangle , \label{sfup} \\
f_{j \downarrow} | \ldots , n_{j \downarrow} , \ldots \rangle &=& (-1)^{\sum_{\ell =1}^L \hat{n}_{\ell \uparrow}} (-1)^{\sum_{\ell < j} \hat{n}_{\ell \downarrow}} | \ldots , n_{j \downarrow}-1 , \ldots \rangle , \label{sfdown}
\end{eqnarray}
where $n_{j \sigma}$ is an eigenvalue of the number operator $\hat{n}_{j \sigma}$.
These operators obey the usual fermionic anticommutation relations including anticommutation of different spins, $\{ f_{j \sigma} , f_{\ell \sigma^{\prime}}^{\dagger} \} = \delta_{j \ell} \delta_{\sigma \sigma^{\prime}}$.

The order of the Jordan-Wigner string factors in Eqs.~(\ref{sfup}) and~(\ref{sfdown}) is unusual. If instead one defines 
annihilation operators $\tilde{f}_{j \sigma}$ for spinful fermions as
\begin{eqnarray}
\tilde{f}_{j \uparrow} | \ldots , n_{j \uparrow} , \ldots \rangle &=&
(-1)^{\sum_{\ell < j} \hat{n}_{\ell \uparrow}}  | \ldots , n_{j \uparrow}-1 , \ldots \rangle , \\
\tilde{f}_{j \downarrow} | \ldots , n_{j \downarrow} , \ldots \rangle &=& (-1)^{\sum_{\ell < j} (\hat{n}_{\ell \uparrow} + \hat{n}_{\ell \downarrow})} (-1)^{\hat{n}_{j \uparrow}} | \ldots , n_{j \downarrow}-1 , \ldots \rangle ,
\end{eqnarray}
then these operators also obey the usual fermionic anticommutation relations including anticommutation of different spins, $\{ {\tilde f}_{j \sigma} , {\tilde f}_{\ell \sigma^{\prime}}^{\dagger} \} = \delta_{j \ell} \delta_{\sigma \sigma^{\prime}}$. They are related to the parafermionic operators as
\begin{eqnarray}
{\tilde f}_{j \uparrow} &=& i^{\sum_{\ell < j} \hat{n}_{\ell\uparrow}} \hat{\Gamma}_j  c_{j} , \label{budefalts} \\
{\tilde f}_{j \downarrow} &=& (-1)^{\hat{n}_{j \uparrow} + \sum_{\ell < j}  \hat{n}_{\ell \downarrow}} c_{j}^2 , \label{bddefalts} \\
c_j &=& (-i)^{\sum_{\ell < j} \hat{n}_{\ell \uparrow}} [ {\tilde f}_{j \uparrow} + (-1)^{\sum_{\ell < j} (\hat{n}_{\ell \uparrow} + \hat{n}_{\ell \downarrow})} {\tilde f}_{j \uparrow}^{\dagger} {\tilde f}_{j \downarrow} ]  . \label{cbdefalts}
\end{eqnarray}

If the matrix ${\cal H}$, Eq.~(14) in the main text, were to be written using the ${\tilde f}_{j \sigma}$ operators, hopping matrix elements would acquire phases dependent on the occupation numbers of the other spin component. In fact, there would be two matrices ${\cal H}_{\sigma}$, where
\begin{eqnarray}
{\cal H}_{\uparrow} &=& \begin{pmatrix}
u_1 & t_1 (-1)^{\hat{n}_{1\downarrow}} & 0 & 0 & \hdots \\
t_1 (-1)^{\hat{n}_{1\downarrow}} & u_2 & t_2 (-1)^{\hat{n}_{2\downarrow}} & 0 & \hdots \\
0 & t_2 (-1)^{\hat{n}_{2\downarrow}} & u_3 & t_3 (-1)^{\hat{n}_{3\downarrow}} & \hdots \\
0 & 0 & t_3 (-1)^{\hat{n}_{3\downarrow}} & u_4 & \hdots \\
\vdots & \vdots & \vdots & \vdots & \ddots
\end{pmatrix} . \\
{\cal H}_{\downarrow} &=& \begin{pmatrix}
u_1 & t_1 (-1)^{\hat{n}_{2\uparrow}} & 0 & 0 & \hdots \\
t_1 (-1)^{\hat{n}_{2\uparrow}} & u_2 & t_2 (-1)^{\hat{n}_{3\uparrow}} & 0 & \hdots \\
0 & t_2 (-1)^{\hat{n}_{3\uparrow}} & u_3 & t_3 (-1)^{\hat{n}_{4\uparrow}} & \hdots \\
0 & 0 & t_3 (-1)^{\hat{n}_{4\uparrow}} & u_4 & \hdots \\
\vdots & \vdots & \vdots & \vdots & \ddots
\end{pmatrix} . 
\end{eqnarray}
However, as is well known for the Su-Schrieffer-Heeger model~\cite{asboth16s}, such phases can be gauged away, leading to the definitions~(\ref{sfup}) and~(\ref{sfdown}), and Eq.~(14) in the main text.

\section{Exact diagonalization of the tight-binding model}

We determine the energy spectrum of the tight-binding model in the main text, Eq.~(4), by exact diagonalization for the simplest nontrivial case of just two orbitals ($L=2$). Specifically, we consider nearest-neighbor hopping $t$ and an onsite energy $u$ with alternating sign,
\begin{eqnarray}
H &=& u (\hat{N}_1 - \hat{N}_2) + t \big[ c_1^{\dagger} (1 - c_1^{\dagger} c_1 + c_1^{\dagger 2} c_1^2) (1 - c_2^{\dagger} c_2 + c_2^{\dagger 2} c_2^2)  c_2 (-i)^{\hat{N}_{1}} + 2 c_1^{\dagger 2} c_2^2 (-1)^{\hat{N}_{1}} + \mathrm{H.c.} \big] .
\end{eqnarray}
The basis has $16$ states, but, owing to number conservation, the Hamiltonian is block diagonal.
In basis $|00\rangle$, $|01\rangle$, $|10\rangle$, $|02\rangle$, $|11\rangle$, $|20\rangle$, $|03\rangle$, $|12\rangle$, $|21\rangle$, $|30\rangle$, $|13\rangle$, $|22\rangle$, $|31\rangle$, $|23\rangle$, $|32\rangle$, $|33\rangle$, the Hamiltonian is
\begin{eqnarray}
\tilde{H} = \begin{pmatrix}
0 & 0 & 0 & 0 & 0 & 0 & 0 & 0 & 0 & 0 & 0 & 0 & 0 & 0 & 0 & 0 \\
0 & -u & t & 0 & 0 & 0 & 0 & 0 & 0 & 0 & 0 & 0 & 0 & 0 & 0 & 0 \\
0 & t & u & 0 & 0 & 0 & 0 & 0 & 0 & 0 & 0 & 0 & 0 & 0 & 0 & 0 \\
0 & 0 & 0 & -2u & 0 & 2t & 0 & 0 & 0 & 0 & 0 & 0 & 0 & 0 & 0 & 0 \\
0 & 0 & 0 & 0 & 0 & 0 & 0 & 0 & 0 & 0 & 0 & 0 & 0 & 0 & 0 & 0 \\
0 & 0 & 0 & 2t & 0 & 2u & 0 & 0 & 0 & 0 & 0 & 0 & 0 & 0 & 0 & 0 \\
0 & 0 & 0 & 0 & 0 & 0 & -3u & t & 2t & 0 & 0 & 0 & 0 & 0 & 0 & 0 \\
0 & 0 & 0 & 0 & 0 & 0 & t & -u & 0 & 2t & 0 & 0 & 0 & 0 & 0 & 0 \\
0 & 0 & 0 & 0 & 0 & 0 & 2t & 0 & u & t & 0 & 0 & 0 & 0 & 0 & 0 \\
0 & 0 & 0 & 0 & 0 & 0 & 0 & 2t & t & 3u & 0 & 0 & 0 & 0 & 0 & 0 \\
0 & 0 & 0 & 0 & 0 & 0 & 0 & 0 & 0 & 0 & -2u & 0 & 2t & 0 & 0 & 0 \\
0 & 0 & 0 & 0 & 0 & 0 & 0 & 0 & 0 & 0 & 0 & 0 & 0 & 0 & 0 & 0 \\
0 & 0 & 0 & 0 & 0 & 0 & 0 & 0 & 0 & 0 & 2t & 0 & 2u & 0 & 0 & 0 \\
0 & 0 & 0 & 0 & 0 & 0 & 0 & 0 & 0 & 0 & 0 & 0 & 0 & -u & t & 0 \\
0 & 0 & 0 & 0 & 0 & 0 & 0 & 0 & 0 & 0 & 0 & 0 & 0 & t & u & 0 \\
0 & 0 & 0 & 0 & 0 & 0 & 0 & 0 & 0 & 0 & 0 & 0 & 0 & 0 & 0 & 0
\end{pmatrix} . \label{edham1}
\end{eqnarray}
The $16$ eigenvalues of $\tilde{H}$ are
\begin{eqnarray}
E &=& \pm 3 \sqrt{u^2 + t^2} , \label{ed1} \\
E &=& \pm 2 \sqrt{u^2 + t^2} , \qquad (\text{twice each}) \nonumber \\
E &=& \pm \sqrt{u^2 + t^2} , \qquad \,\, (\text{three times each}) \nonumber \\
E &=& 0 . \qquad\qquad\qquad\quad (\text{four times}) \nonumber 
\end{eqnarray}

The single-particle spectrum is obtained by diagonalizing the matrix ${\cal H}$, Eq.~(14) in the main text, which, for $L=2$, is
\begin{eqnarray}
{\cal H} = \begin{pmatrix}
u & t \\
t & -u
\end{pmatrix} .
\end{eqnarray}
This has two eigenvalues, $\epsilon_1 = \sqrt{u^2 + t^2}$ and $\epsilon_2 = -\sqrt{u^2 + t^2}$.
According to Eq.~(2) in the main text, this describes the energy spectrum as
\begin{eqnarray}
E = n_1 \epsilon_1 + n_2 \epsilon_2 , \label{smsp}
\end{eqnarray}
where $n_1 = 0,1,2,3$ and $n_2 = 0,1,2,3$, in agreement with the exact diagonalization of $H$, Eq.~(\ref{ed1}).

\section{Model with next-nearest-neighbor hopping}

Here we construct a Hamiltonian with a single-particle spectrum [Eq.~(2) in the main text] written in terms of parafermionic operators and including next-nearest-neighbor hopping.
We begin with a next-nearest-neighbor hopping term for fermions,
\begin{eqnarray}
\delta H = \sum_{j=1}^{L-2} v_j f_j^{\dagger} f_{j+2} + \mathrm{H.c.} ,
\end{eqnarray}
where $v_j$ is the hopping parameter from site $j$ to site $j+2$, and we consider open boundary conditions.
Using the mapping, Eqs.~(7) and (8) in the main text, the Hamiltonian may be written in terms of parafermion operators as
\begin{eqnarray}
H &=& \sum_{j=1}^L u_j \hat{N}_j + \sum_{j=1}^{L-1} t_j \big[ c_j^{\dagger} \hat{\Gamma}_j \hat{\Gamma}_{j+1}  c_{j+1} (-i)^{\hat{N}_{j}} 
+ 2 c_j^{\dagger 2} c_{j+1}^2 (-1)^{\hat{N}_{j}} + \mathrm{H.c.} \big] \nonumber \\
&& + \sum_{j=1}^{L-2} v_j \big[ c_j^{\dagger} \hat{\Gamma}_j \hat{\Gamma}_{j+2}  c_{j+2} i^{(\hat{N}_{j} + \hat{N}_{j+1})} 
+ 2 c_j^{\dagger 2} c_{j+2}^2 (-1)^{(\hat{n}_{j \uparrow}+\hat{n}_{j \downarrow}+\hat{n}_{j+1 \uparrow}+\hat{n}_{j+1 \downarrow})} + \mathrm{H.c.} \big] ,
\end{eqnarray}
where $\hat{\Gamma}_j = 1 - c_j^{\dagger} c_j + c_j^{\dagger 2} c_j^2$. Here we have included the contribution of onsite energies and nearest-neighbor hopping [Eq.~(4) in the main text].
The mapping means that the single-particle energy levels $\epsilon_{\ell}$ of $H$, with $\ell = 1,2,\ldots , L$, are given by the eigenvalues of an $L \times L$ matrix ${\cal H}$,
\begin{eqnarray}
{\cal H} = \begin{pmatrix}
u_1 & t_1 & v_1 & 0 & \hdots \\
t_1 & u_2 & t_2 & v_2 & \hdots \\
v_1 & t_2 & u_3 & t_3 & \hdots \\
0 & v_2 & t_3 & u_4 & \hdots \\
\vdots & \vdots & \vdots & \vdots & \ddots
\end{pmatrix} .
\end{eqnarray}

\section{Kitaev superconducting chain}\label{s:kitaev}

As a more complicated example, we consider the 1D Kitaev superconducting chain~\cite{kitaev01s} which is a mean-field model of a topological superconductor including superconducting pairing terms as well as noninteracting hopping terms.
The counterpart parafermion model with open boundary conditions is
\begin{eqnarray}
H &=& -\mu \sum_{j=1}^L \hat{N}_j + t \sum_{j=1}^{L-1} \big[ c_j^{\dagger} \hat{\Gamma}_j \hat{\Gamma}_{j+1}  c_{j+1} (-i)^{\hat{N}_{j}} + 2 c_j^{\dagger 2} c_{j+1}^2 (-1)^{\hat{N}_{j}} + \mathrm{H.c.} \big] \nonumber \\
&& \qquad + \Delta \sum_{j=1}^{L-1} \big[ c_j^{\dagger} \hat{\Gamma}_j c_{j+1}^{\dagger} \hat{\Gamma}_{j+1} (-1)^{\sum_{\ell < j} \hat{N}_{\ell}} i^{\hat{N}_j}
+ 2 c_j^{\dagger 2} c_{j+1}^{\dagger 2} (-1)^{\hat{N}_{j}}
+ \mathrm{H.c.} \big] , \label{hampk}
\end{eqnarray}
where
\begin{eqnarray}
\hat{\Gamma}_j = 1 - c_j^{\dagger} c_j + c_j^{\dagger 2} c_j^2 . 
\end{eqnarray}
Here $\mu$ is the chemical potential, $t$ is the parameter for nearest-neighbor hopping, and $\Delta$ is the pairing term, and there is translational invariance with one orbital per cell. Without loss of generality, we assume that $\Delta$ is real.
Using the mapping~(7) to~(9) in the main text, the parafermionic Hamiltonian is $H = H_{\uparrow} + 2 H_{\downarrow}$, where the fermionic models are
\begin{eqnarray}
H_{\sigma} &=& - \mu \sum_{j=1}^L f_{j \sigma}^{\dagger} f_{j \sigma}
+ t \sum_{j=1}^{L-1} ( f_{j \sigma}^{\dagger} f_{j+1 \sigma} + \mathrm{H.c.} ) + \Delta \sum_{j=1}^{L-1} ( f_{j \sigma}^{\dagger} f_{j+1 \sigma}^{\dagger} + \mathrm{H.c.} ) . \label{hsigmak}
\end{eqnarray}
This may be solved by writing $H_{\sigma}$ in the Bogoliubov de Gennes (BdG) representation~\cite{alicea12s,leijnse12s,beenakker15s,guo16s,sato17s},
\begin{eqnarray}
H_{\sigma} = \frac{1}{2} \Phi_{\sigma}^{\dagger} {\cal H} \Phi_{\sigma} - \frac{1}{2} \mu L  , 
\end{eqnarray}
where ${\cal H}$ is a $2L \times 2L$ matrix.
We write this explicitly for $L=4$ sites with
\begin{eqnarray}
\Phi_{\sigma}^{\dagger} = \begin{pmatrix}
f_{1 \sigma}^{\dagger} & f_{1 \sigma} & f_{2 \sigma}^{\dagger} & f_{2 \sigma} & f_{3 \sigma}^{\dagger} & f_{3 \sigma} & f_{4 \sigma}^{\dagger} & f_{4 \sigma}
\end{pmatrix} ,
\end{eqnarray}
then the BdG matrix is
\begin{eqnarray}
{\cal H} = \begin{pmatrix}
-\mu & 0 & t & \Delta & 0 & 0 & 0 & 0 \\
0 & \mu & -\Delta & -t & 0 & 0 & 0 & 0 \\
t & -\Delta & -\mu & 0 & t & \Delta & 0 & 0 \\
\Delta & -t & 0 & \mu & -\Delta & -t & 0 & 0 \\
0 & 0 & t & -\Delta & -\mu & 0 & t & \Delta \\
0 & 0 & \Delta & -t & 0 & \mu & -\Delta & -t \\
0 & 0 & 0 & 0 & t & -\Delta & -\mu & 0 \\
0 & 0 & 0 & 0 & \Delta & -t & 0 & \mu
\end{pmatrix} .
\end{eqnarray}
Denoting the positive energy eigenvalues of ${\cal H}$ as $\epsilon_{\ell}$ for $\ell = 1,2,\ldots, L$, then the parafermionic ground state energy $E_0 = E_{0 \uparrow} + 2 E_{0 \downarrow}$ is
\begin{eqnarray}
E_0 = - \frac{3}{2} \Big(   \mu L + \sum_{\ell} \epsilon_{\ell} \Big) . \label{gs}
\end{eqnarray}
The many-body energy spectrum of the Hamiltonian~(\ref{hampk}) consists of $4^L$ energies, each of which may be expressed as
\begin{eqnarray}
E = E_0 + \sum_{{\ell}=1}^L n_{\ell} \epsilon_{\ell} , \label{pkspectrum}
\end{eqnarray}
where $n_{\ell} = 0,1,2,3$ are parafermion occupation numbers.

For simplicity, we consider the topological phase in the dimer limit $\Delta = t$ and $\mu = 0$, in which case one of the (nominally positive) single-particle levels is zero giving a four-fold degeneracy of the ground state, according to Eq.~(\ref{pkspectrum}).
The Hamiltonian~(\ref{hampk}) preserves the parity of the system so that the total number of parafermions $N_{\mathrm{tot}}$ is either an even ($0$) or odd ($1$) number. Owing to the mapping to the spin-$1/2$ fermions, the parity of the total numbers of spin up fermions $n_{\mathrm{tot} \uparrow}$ is also conserved, as is the parity of the total numbers of spin down fermions $n_{\mathrm{tot} \downarrow}$. With the condition $N_{\mathrm{tot}} = n_{\mathrm{tot} \uparrow} + 2 n_{\mathrm{tot} \downarrow}$, the four degenerate ground states may be labeled with the parities of $n_{\mathrm{tot} \uparrow}$ and $n_{\mathrm{tot} \downarrow}$ as $(n_{\mathrm{tot} \uparrow} , n_{\mathrm{tot} \downarrow}) = \{ (0,0) , (1,0) , (0,1) , (1,1) \}$, where $n_{\mathrm{tot} \uparrow}$ has the same parity as $N_{\mathrm{tot}}$.

Baxter's non-Hermitian clock model~\cite{baxter89as,baxter89bs,fendley14s,alicea16s,alcaraz17s,batchelor23s} may be viewed as a natural parafermionic generalization of the Kitaev superconducting chain~\cite{kitaev01s} because, particularly, it can be conveniently represented using Weyl parafermions~\cite{fendley14s,alicea16s,cobanera14s} which are generalizations of fermionic Majorana operators~\cite{alicea12s,leijnse12s,beenakker15s}.
Although the Hamiltonian~(\ref{hampk}) is also related to the Kitaev chain~\cite{kitaev01s}, its representation in terms of Weyl parafermions is not particularly useful. Instead, we consider the fermionic spin-$1/2$ Hamiltonian $H_{\sigma}$ where $H = H_{\uparrow} + 2 H_{\downarrow}$.
In the spin-$1/2$ representation, Majorana operators~\cite{alicea12s,leijnse12s,beenakker15s} are defined as
\begin{eqnarray*}
\gamma_{jA\sigma} &=& f_{j \sigma} + f_{j \sigma}^{\dagger} ; \qquad \qquad \,\;
\gamma_{jB\sigma} = i (f_{j \sigma} - f_{j \sigma}^{\dagger}) , \\
f_{j \sigma} &=& \frac{1}{2} (\gamma_{jA\sigma} - i \gamma_{jB\sigma}) ; \qquad
f_{j \sigma}^{\dagger} = \frac{1}{2} (\gamma_{jA\sigma} + i \gamma_{jB\sigma}) ,
\end{eqnarray*}
where $\gamma_{j\alpha\sigma}^{\dagger} = \gamma_{j\alpha\sigma}$ and $\{ \gamma_{j\alpha\sigma} \gamma_{j^{\prime}\alpha^{\prime}\sigma} \} = 2 \delta_{jj^{\prime}} \delta_{\alpha\alpha^{\prime}}$.
Then the spin-$1/2$ Hamiltonian~(\ref{hsigmak}) in the topological dimer limit ($\Delta = t$ and $\mu = 0$) is
\begin{eqnarray}
H_{\sigma} = i t \sum_{j=1}^{L-1} \gamma_{jB\sigma} \gamma_{j+1A\sigma} .
\end{eqnarray}
The edge Majorana operators $\gamma_{1A\sigma}$ and $\gamma_{LB\sigma}$ are absent from this Hamiltonian, and they may be combined into a zero-energy fermionic annihilation operator as $f_{\mathrm{z}\sigma} = (\gamma_{1A\sigma} - i \gamma_{LB\sigma}) / 2$.
The parity of the number of spin-$1/2$ fermions corresponds to the eigenvalue of the number operator $\hat{n}_{\mathrm{z} \sigma} = f_{\mathrm{z}\sigma}^{\dagger} f_{\mathrm{z}\sigma}$ with $0$~($1$) eigenvalue for even (odd) parity~\cite{leijnse12s}.

Using the mapping~(7) and~(8) in the main text, the two zero-energy fermionic operators may be represented as
\begin{eqnarray}
f_{\mathrm{z}\uparrow} &=& \frac{1}{2} \big[ \hat{\Gamma}_1 c_1 + c_1^{\dagger} \hat{\Gamma}_1 + i^{\sum_{\ell < L} \hat{N}_{\ell}} \hat{\Gamma}_L c_L - (-i)^{\sum_{\ell < L} \hat{N}_{\ell}} c_L^{\dagger} \hat{\Gamma}_L \big] , \\
f_{\mathrm{z}\uparrow} &=& \frac{1}{2} \big[ c_1^2 + c_1^{\dagger 2}
+ (-1)^{\sum_{\ell < L} (\hat{n}_{\ell \uparrow} + \hat{n}_{\ell \downarrow})}
( c_L^2 - c_L^{\dagger 2} ) \big] .
\end{eqnarray}
In the dimer limit $\Delta = t$ and $\mu = 0$, the occupation of the zero energy modes for the Hamiltonian~(\ref{hampk}) is given by the parafermion operator $\hat{N}_{\mathrm{z}} = \hat{n}_{\mathrm{z} \uparrow} + 2 \hat{n}_{\mathrm{z} \downarrow} = f_{\mathrm{z}\uparrow}^{\dagger} f_{\mathrm{z}\uparrow} + 2 f_{\mathrm{z}\downarrow}^{\dagger} f_{\mathrm{z}\downarrow}$, and its eigenvalues $0,1,2,3$ label the degenerate ground states with
$(n_{\mathrm{z} \uparrow} , n_{\mathrm{z} \downarrow}) = \{ (0,0) , (1,0) , (0,1) , (1,1) \}$.
With two decoupled Kitaev chains $H = H_{\uparrow} + 2 H_{\downarrow}$, this system corresponds to a `weakly interacting' parafermion phase~\cite{calzona18s,teixeira21s} as indicated, for instance, by the weight of the peak in the fermionic local spectral function at the edges of the system, which is double that of a strongly correlated parafermion phase~\cite{calzona18s}.

\section{Mapping for $p=6$}

We begin by defining creation and annihilation operators for $p$-state parafermions~\cite{cobanera14s}.
We use
\begin{eqnarray}
\omega = \exp (2 \pi i / p) ; \qquad \qquad
\bar{\omega} = \exp (-2 \pi i / p) . \label{gens}
\end{eqnarray}
Parafermion creation $c_j^{\dagger}$ and annihilation $c_j$ operators act on Fock states~\cite{cobanera14s} as
\begin{eqnarray*}
c_j^{\dagger} | n_1 , \ldots , n_j , \ldots n_L \rangle &=&
\bar{\omega}^{\sum_{\ell < j} n_{\ell}}  | n_1 , \ldots , n_j + 1 , \ldots n_L \rangle , \\
c_j | n_1 , \ldots , n_j , \ldots n_L \rangle &=&
\omega^{\sum_{\ell < j} n_{\ell}}  | n_1 , \ldots , n_j - 1 , \ldots n_L \rangle ,
\end{eqnarray*}
where $n_j = 0,1,\ldots , p-1$ is an eigenvalue of the number operator $\hat{N}_j$~\cite{cobanera14s},
\begin{eqnarray}
\hat{N}_j | n_1 , \ldots , n_j , \ldots n_L \rangle =
n_j | n_1 , \ldots , n_j , \ldots n_L \rangle , \label{pn1s}
\end{eqnarray}
which is given by
\begin{eqnarray}
\hat{N}_j = \sum_{m=1}^{p-1} c_j^{\dagger m} c_j^m . \label{ngen}
\end{eqnarray}
Creation and annihilation operators have the following commutation relations~\cite{cobanera14s},
\begin{eqnarray}
c_j^{\dagger p} &=& 0 , \qquad
c_j^{\dagger} c_{\ell}^{\dagger} = \omega c_{\ell}^{\dagger} c_j^{\dagger}
\quad \text{for}\,\, j < \ell , \\
c_j^p &=& 0 , \qquad
c_j c_{\ell} = \omega c_{\ell} c_j
\quad \text{for}\,\, j < \ell , \\
c_j^{\dagger} c_{\ell} &=& \bar{\omega} c_{\ell} c_j^{\dagger} , \qquad
c_j c_{\ell}^{\dagger} = \bar{\omega} c_{\ell}^{\dagger} c_j
\quad \text{for}\,\, j < \ell .
\end{eqnarray}
For $c_j^{\dagger}$ and $c_j$ on the same orbital, there are $p-1$ relations:
\begin{eqnarray}
c_j^{\dagger m} c_j^m + c_j^{p-m} c_j^{\dagger (p-m)} = 1 , \label{genf}
\end{eqnarray}
for $m = 1 , 2 , \ldots p-1$.

For $p=6$, parafermions may be mapped onto a species of $p=3$ parafermions and a species of fermions. We denote the $p=3$ parafermion annihilation operator on site $j$ as $g_{jy}$ and the fermion annihilation operator on site $j$ as $f_{jx}$.
In terms of the $p=6$ parafermion annihilation operator $c_j$, they may be written as
\begin{eqnarray}
f_{jx} &=&  \bar{\omega}^{\sum_{\ell < j} \hat{N}_{\ell}} (-1)^{\sum_{\ell < j} \hat{n}_{\ell x}}\hat{\Gamma}_j  c_j , \\
g_{jy} &=& \bar{\omega}^{\sum_{\ell < j} 2 (\hat{n}_{\ell x} + \hat{n}_{\ell y})} c_{j}^2 ,
\end{eqnarray}
where $\omega = \exp (\pi i / 3)$, $\bar{\omega} = \exp (-\pi i / 3)$ and
\begin{eqnarray}
\hat{\Gamma}_j &=& \sum_{m=0}^{4} (-1)^{m} c_j^{\dagger m} c_j^m , \\
\hat{n}_{j x} &=& f_{j x}^{\dagger} f_{j x} , \\
\hat{n}_{j y} &=& g_{j y}^{\dagger} g_{j y} + g_{j y}^{\dagger 2} g_{j y}^2 .
\end{eqnarray}
The annihilation operator for $p=6$ parafermions may be written as
\begin{eqnarray}
c_j = \alpha_j f_{j x} + \beta_j f_{j x}^{\dagger} g_{j y} ,
\end{eqnarray}
where
\begin{eqnarray}
\alpha_j &=& (-1)^{\sum_{\ell < j} \hat{n}_{\ell x}} \omega^{\sum_{\ell < j} \hat{N}_{\ell}} , \\
\beta_j &=& (-\omega)^{\sum_{\ell < j} \hat{n}_{\ell x}} .
\end{eqnarray}
For simplicity, we choose a gauge so that operators for different species commute, $[ f_{j x} , g_{\ell y}^{\dagger} ] = 0$, etc. The parafermionic number operator~(\ref{ngen}) may be expressed as
\begin{eqnarray}
\hat{N}_j = \hat{n}_{\ell x} + 2 \hat{n}_{\ell y} = f_{j x}^{\dagger} f_{j x} + 2 g_{j y}^{\dagger} g_{j y} + 2 g_{j y}^{\dagger 2} g_{j y}^2 .
\end{eqnarray}

\section{Mapping for $p=8$}

We consider the parafermion creation and annihilation operators,
Eqs.~(\ref{gens})-(\ref{genf}), for $p=8$.
For $p=8$, parafermions may be mapped onto three different species of fermions, which we denote with labels $x$, $y$, and $z$.
The annihilation operators $f_{j \sigma}$, $\sigma = x,y,z$, for spinful fermions act on Fock states with $n_{j \sigma} = 1$ as
\begin{eqnarray}
f_{j x} | \ldots , n_{j x} , \ldots \rangle &=&
(-1)^{\sum_{\ell < j} \hat{n}_{\ell x}}  | \ldots , n_{j x}-1 , \ldots \rangle , \label{sfx} \\
f_{j y} | \ldots , n_{j y} , \ldots \rangle &=& (-1)^{\sum_{\ell =1}^L \hat{n}_{\ell x}} (-1)^{\sum_{\ell < j} \hat{n}_{\ell y}} | \ldots , n_{j y}-1 , \ldots \rangle , \label{sfy} \\
f_{j z} | \ldots , n_{j z} , \ldots \rangle &=& (-1)^{\sum_{\ell =1}^L ( \hat{n}_{\ell x} + \hat{n}_{\ell y})} (-1)^{\sum_{\ell < j} \hat{n}_{\ell z}} | \ldots , n_{j z}-1 , \ldots \rangle , \label{sfz}
\end{eqnarray}
where $n_{j \sigma}$ is an eigenvalue of the number operator $\hat{n}_{j \sigma}$.
These operators obey the usual fermionic anticommutation relations including anticommutation of different spins, $\{ f_{j \sigma} , f_{\ell \sigma^{\prime}}^{\dagger} \} = \delta_{j \ell} \delta_{\sigma \sigma^{\prime}}$.

For the mapping to $p=8$ parafermion annihilation operators, $c_j$, the fermion annihilation operators for site $j$ may be written as
\begin{eqnarray}
f_{j x} &=& \bar{\omega}^{\sum_{\ell < j} \hat{N}_{\ell}} (-1)^{\sum_{\ell < j} \hat{n}_{\ell x}} \hat{\Gamma}_j  c_j , \\
f_{j y} &=& (-1)^{\sum_{\ell = 1}^L \hat{n}_{\ell x}} (-i)^{\sum_{\ell < j} \hat{n}_{\ell x}} \hat{\Lambda}_j  c_j^2 , \\
f_{j z} &=& (-1)^{\sum_{\ell = 1}^L (\hat{n}_{\ell x} + \hat{n}_{\ell y})} (-1)^{\sum_{\ell < j} (\hat{n}_{\ell x} + \hat{n}_{\ell z})} c_{j}^4 ,
\end{eqnarray}
where $\omega = \exp (\pi i / 4)$, $\bar{\omega} = \exp (-\pi i / 4)$, and
\begin{eqnarray}
\hat{\Gamma}_j &=& \sum_{m=0}^{6} (-1)^{m} c_j^{\dagger m} c_j^m , \\
\hat{\Lambda}_j &=& 1 - c_j^{\dagger 2} c_j^2 + c_j^{\dagger 4} c_j^4 , \\
\hat{n}_{j x} &=& f_{j x}^{\dagger} f_{j x} , \\
\hat{n}_{j y} &=& f_{j y}^{\dagger} f_{j y} , \\
\hat{n}_{j z} &=& f_{j z}^{\dagger} f_{j z} .
\end{eqnarray}
The annihilation operator for $p=8$ parafermions may be written as
\begin{eqnarray}
c_j = \alpha_j f_{j x} + \beta_j f_{j x}^{\dagger} f_{j y} + \gamma_j f_{j x}^{\dagger} f_{j y}^{\dagger} f_{j z} ,
\end{eqnarray}
where
\begin{eqnarray}
\alpha_j &=& (-1)^{\sum_{\ell < j} \hat{n}_{\ell x}} \omega^{\sum_{\ell < j} \hat{N}_{\ell}} , \\
\beta_j &=& (-1)^{\sum_{\ell > j} \hat{n}_{\ell x}} (-1)^{\sum_{\ell < j} \hat{n}_{\ell y}} \omega^{\sum_{\ell < j} \hat{N}_{\ell}} , \\
\gamma_j &=& (-1)^{\sum_{\ell > j} \hat{n}_{\ell y}} (-1)^{\sum_{\ell < j} (\hat{n}_{\ell x} + \hat{n}_{\ell z})} \omega^{\sum_{\ell < j} \hat{N}_{\ell}} .
\end{eqnarray}
The parafermionic number operator~(\ref{ngen}) may be expressed as
\begin{eqnarray}
\hat{N}_j = f_{j x}^{\dagger} f_{j x} + 2 f_{j y}^{\dagger} f_{j y} + 4 f_{j z}^{\dagger} f_{j z} .
\end{eqnarray}


\begin{thebibliography}{99}

\bibitem{fendley14}
P. Fendley,
Free parafermions,
J. Phys. A: Math. Theor. {\bf 47}, 075001 (2014).

\bibitem{fendley24}
P. Fendley and B. Pozsgay,
Free fermions beyond Jordan and Wigner,
SciPost Phys. {\bf 16}, 102 (2024).

\bibitem{cobanera14}
E. Cobanera and G. Ortiz,
Fock parafermions and self-dual representations of the braid group,
Phys.\ Rev.\ A {\bf 89}, 012328 (2014).

\bibitem{alicea16}
J. Alicea and P. Fendley,
Topological Phases with Parafermions: Theory and Blueprints,
Annu.\ Rev.\ Condens.\ Matter Phys.\ {\bf 7}, 119 (2016).

\bibitem{hutter16}
A. Hutter and D. Loss,
Quantum computing with parafermions,
Phys.\ Rev.\ B {\bf 93}, 125105 (2016).

\bibitem{kitaev03}
A. Y. Kitaev,
Fault-tolerant quantum computation by anyons,
Ann.\ Phys.\ {\bf 303}, 2 (2003).

\bibitem{nayak08}
C. Nayak, S. H. Simon, A. Stern, M. Freedman, and S. Das Sarma,
Non-Abelian anyons and topological quantum computation,
Rev.\ Mod.\ Phys.\ {\bf 80}, 1083 (2008).

\bibitem{lahtinen17}
V. T. Lahtinen and J. K. Pachos,
A Short Introduction to Topological Quantum Computation,
SciPost Phys.\ {\bf 3}, 021 (2017).

\bibitem{fendley12}
P. Fendley,
Parafermionic edge zero modes in $Z_n$-invariant spin chains,
J.\ Stat.\ Mech.\ {\bf 1211}, P11020 (2012).

\bibitem{nigro14}
A. Nigro and M. Gherardi,
A parafermionic generalization of the Jaynes–Cummings model,
J.\ Phys.\ A: Math.\ Theor.\ {\bf 47}, 265205 (2014).

\bibitem{milsted14}
A. Milsted, E. Cobanera, M. Burrello, and G. Ortiz,
Commensurate and incommensurate states of topological quantum matter,
Phys.\ Rev.\ B {\bf 90}, 195101 (2014).

\bibitem{jermyn14}
A. S. Jermyn, R. S. K. Mong, J. Alicea1, and P. Fendley,
Stability of zero modes in parafermion chains,
Phys.\ Rev.\ B {\bf 90}, 165106 (2014).

\bibitem{zhuang15}
Y. Zhuang, H. J. Changlani, N. M. Tubman, and T. L. Hughes,
Phase diagram of the $Z_3$ parafermionic chain with chiral interactions,
Phys.\ Rev.\ B {\bf 92}, 035154 (2015).

\bibitem{sreejith16}
G. J. Sreejith, A. Lazarides and R. Moessner,
Parafermion chain with $2\pi / k$ Floquet edge modes,
Phys.\ Rev.\ B {\bf 94}, 045127 (2016).

\bibitem{iemini17}
F. Iemini, C. Mora, and L. Mazza,
Topological Phases of Parafermions: A Model with Exactly Solvable Ground States,
Phys.\ Rev.\ Lett.\ {\bf 118}, 170402 (2017).

\bibitem{samajdar18}
R. Samajdar, S. Choi, H. Pichler, M. D. Lukin, and S. Sachdev,
Numerical study of the chiral $Z_3$ quantum phase transition in one spatial dimension,
Phys.\ Rev.\ A {\bf 98}, 023614 (2018).

\bibitem{zhang19}
S.-Y. Zhang, H.-Z. Xu, Y.-X. Huang, G.-C. Guo, Z.-W. Zhou, and M. Gong,
Topological phase, supercritical point, and emergent phenomena in an extended parafermion chain,
Phys.\ Rev.\ B {\bf 100}, 125101 (2019).

\bibitem{rossini19}
D. Rossini, M. Carrega, M. Calvanese Strinati, and L. Mazza,
Anyonic tight-binding models of parafermions and of fractionalized fermions,
Phys.\ Rev.\ B {\bf 99}, 085113 (2019).

\bibitem{schmidt20}
T. L. Schmidt,
Bosonization for fermions and parafermions,
Eur.\ Phys.\ J.\ Special Topics {\bf 229}, 621 (2020).

\bibitem{offeidanso20}
A. Offei-Danso, F. M. Surace, F. Iemini, A. Russomanno, and R. Fazio,
Quantum clock models with infinite-range interactions,
J.\ Stat.\ Mech.\ 073107 (2020).

\bibitem{lahtinen21}
V. Lahtinen, T. Mansson, and E. Ardonne,
Quantum criticality in many-body parafermion chains,
SciPost Phys.\ Core {\bf 4}, 014 (2021).

\bibitem{teixeira22b}
R. L. R. C. Teixeira,  A. Haller, R. Singh, A. Mathew, E. G. Idrisov, L. G. G. V. Dias da Silva, and T. L. Schmidt,
Overlap of parafermionic zero modes at a finite distance,
Phys.\ Rev.\ Research {\bf 4}, 043094 (2022).

\bibitem{baxter89a}
R. J. Baxter,
A simple solvable $Z_N$ Hamiltonian,
Phys. Lett. A {\bf 140}, 155 (1989).

\bibitem{baxter89b}
R. J. Baxter,
Superintegrable chiral Potts model: Thermodynamic properties, an “inverse” model, and a simple associated Hamiltonian.
J. Stat. Phys. {\bf 57}, 1 (1989).

\bibitem{alcaraz17}
F. C. Alcaraz, M. T. Batchelor, and Z.-Z. Liu,
Energy spectrum and critical exponents of the free parafermion $Z_N$ spin chain,
J.\ Phys.\ A: Math.\ Theor.\ {\bf 50}, 16LT03 (2017).

\bibitem{alcaraz20a}
F. C. Alcaraz and R. A. Pimenta,
Free fermionic and parafermionic quantum spin chains with multispin interactions,
Phys.\ Rev.\ B {\bf 102}, 121101(R) (2020).

\bibitem{alcaraz20b}
F. C. Alcaraz and R. A. Pimenta,
Integrable quantum spin chains with free fermionic and parafermionic spectrum,
Phys.\ Rev.\ B {\bf 102},  235170 (2020).

\bibitem{alcaraz21}
F. C. Alcaraz and R. A. Pimenta,
Free-parafermionic $Z(N)$ and free-fermionic $XY$ quantum chains,
Phys.\ Rev.\ E {\bf 104}, 054121 (2021).

\bibitem{mastiukova22}
A. S. Mastiukova, D. V. Kurlov, V. Gritsev, and A. K. Fedorov,
Free Fock Parafermions in the Tight-binding Model with Dissipation,
JETP Lett.\ {\bf 123}, 359 (2026).

\bibitem{batchelor23}
M. T. Batchelor, R. A. Henry, and X. Lu,
A brief history of free parafermions,
AAPPS Bulletin {\bf 33}, 29  (2023).

\bibitem{mann25}
R. L. Mann, S. J. Elman, D. R. Wood and A. Chapman,
A graph-theoretic framework for free-parafermion solvability,
Proc.\ R.\ Soc.\ A {\bf 481}, 20240671 (2025).

\bibitem{orbitalcomment}
Baxter's clock model~\cite{baxter89a,baxter89b,fendley14,alicea16} consists of a spin chain where each site consists of a $p$-state spin.
In the representation of indistinguishable particles with creation and annihilation operators, each site consists of a single state occupied by $0,1 , \ldots ,p-1$ particles. In this paper, we refer to this state as an `orbital', by analogy with atomic orbitals for electrons.
The spin chain may be written in terms of the indistinguishable particles through the Fradkin-Kadanoff mapping~\cite{fradkin80,alcaraz81,cobanera14}.

\bibitem{fradkin80}
E. Fradkin and L. P. Kadanoff,
Disorder variables and para-fermions in two-dimensional statistical mechanics,
Nucl.\ Phys.\ B {\bf 170}, 1 (1980).

\bibitem{alcaraz81}
F. C. Alcaraz and R. K\"oberle,
Hidden parafermions in Z(N) theories,
Phys.\ Rev.\ D {\bf 24}, 1562 (1981).

\bibitem{gentile40}
G. Gentile,
Osservazioni sopra le statistiche intermedie,
Nuovo Cimento {\bf 17}, 493 (1940).

\bibitem{gentile42}
G. Gentile,
Le statistiche intermedie e le proprietà dell’elio liquido,
Nuovo Cimento {\bf 19}, 109 (1942).

\bibitem{asboth16}
J. K. Asb\'oth, L. Oroszl\'any, and A. P\'alyi,
{\it A Short Course on Topological Insulators}
(Springer, Switzerland, 2016).

\bibitem{mccann23}
E. McCann,
Catalog of noninteracting tight-binding models with two energy bands in one dimension,
Phys.\ Rev.\ B {\bf 107}, 245401 (2023).

\bibitem{xu17}
W.-T. Xu and G.-M. Zhang,
Matrix product states for topological phases with parafermions,
Phys.\ Rev.\ B {\bf 95}, 195122 (2017).

\bibitem{calzona18}
A. Calzona, T. Meng, M. Sassetti, and T. L. Schmidt,
$\mathbb{Z}_4$ parafermions in one-dimensional fermionic lattices,
Phys.\ Rev.\ B {\bf 98}, 201110(R) (2018).

\bibitem{mahyaeh20}
I. Mahyaeh, J. Wouters, and D. Schuricht,
Phase diagram of the $\mathbb{Z}_3$-Fock parafermion chain with pair hopping,
SciPost Phys.\ Core {\bf 3}, 011 (2020).

\bibitem{camacho22}
G. Camacho, J. Vahedi, D. Schuricht, and C. Karrasch,
Disorder effects in the $\mathbb{Z}_3$ Fock parafermion chain,
Phys.\ Rev.\ B {\bf 106}, 235132 (2022).

\bibitem{bahovadinov22}
M. S. Bahovadinov, W. Buijsman, A. K. Fedorov,  V. Gritsev, and D. V. Kurlov,
Many-body localization of $\mathbb{Z}_3$ Fock parafermions,
Phys.\ Rev.\ B {\bf 106}, 224205 (2022).

\bibitem{kohmoto81}
M. Kohmoto, M. den Nijs, and L. P. Kadanoff,
Hamiltonian studies of the $d=2$ Ashkin-Teller model,
Phys.\ Rev.\ B {\bf 24}, 5229 (1981).

\bibitem{yamanaka94}
M. Yamanaka, Y. Hatsugai, and M. Kohmoto,
Phase diagram of the Ashkin-Teller quantum spin chain,
Phys. Rev. B {\bf 50}, 559 (1994).

\bibitem{hutter15}
A. Hutter, J. R. Wootton, and D. Loss,
Parafermions in a Kagome Lattice of Qubits for Topological Quantum Computation
Phys.\ Rev.\ X {\bf 5}, 041040 (2015).

\bibitem{yu15}
L.-W. Yu and M.-L. Ge,
$\mathbb{Z}_3$ parafermionic chain emerging from Yang-Baxter equation,
Sci.\ Rep.\ {\bf 6}, 21497 (2016).

\bibitem{meichanetzidis18}
K. Meichanetzidis, C. J. Turner, A. Farjami, Z. Papi'c, and J. K. Pachos,
Free-fermion descriptions of parafermion chains and string-net models,
Phys.\ Rev.\ B {\bf 97}, 125104 (2018).

\bibitem{chew18}
A. Chew, D. F. Mross, and J. Alicea,
Fermionized parafermions and symmetry-enriched Majorana modes,
Phys.\ Rev.\ B {\bf 98}, 085143 (2018).

\bibitem{bomantara21}
R. W. Bomantara,
$\mathbb{Z}_4$ parafermion $\pm \pi/2$ modes in an interacting periodically driven superconducting chain,
Phys.\ Rev.\ B {\bf 104}, L121410 (2021).

\bibitem{teixeira22}
R. L. R. C. Teixeira and L. G. G. V. Dias da Silva,
Edge $\mathbb{Z}_3$ parafermions in fermionic lattices,
Phys.\ Rev.\ B {\bf 105}, 195121 (2022).

\bibitem{traverso23}
S. Traverso, C. Fleckenstein, M. Sassetti, N. T. Ziani,
An exact local mapping from clock-spins to fermions,
SciPost Phys.\ Core {\bf 6}, 055 (2023).

\bibitem{osvath24}
B. Osv\'ath, G. Barcza, \"O. Legeza, B. D\'ora, and L. Oroszlány,
Electronic ladder model harboring $\mathbb{Z}_4$ parafermions,
Phys.\ Rev.\ B {\bf 110}, 085304 (2024).

\bibitem{safwan25}
A. H. Safwan and R. W. Bomantara,
Generating non-Clifford gate operations through exact mapping between Majorana fermions and $\mathbb{Z}_4$ parafermions,
J.\ Phys.\ A: Math.\ Theor.\ {\bf 58}, 335302 (2025).

\bibitem{bondesan13}
R. Bondesan and T. Quella,
Topological and symmetry broken phases of $Z_N$ parafermions in one dimension,
J. Stat. Mech. P10024 (2013).

\bibitem{moran17}
N. Moran, D. Pellegrino, J. K. Slingerland, and G. Kells,
Parafermionic clock models and quantum resonance,
Phys.\ Rev.\ B {\bf 95}, 235127 (2017).

\bibitem{supplementary}
The Supplemental Material (see also references~\cite{alicea12,leijnse12,beenakker15,guo16,sato17,teixeira21} therein) gives more details about the mapping from $p=4$ parafermions to spinful fermions. It includes a description of exact diagonalization of the tight-binding model~(\ref{ham1}) for the simplest nontrivial case of two orbitals. There are also two further examples of solvable models for $p=4$:
The first includes longer-range hopping giving a nonlocal parafermion Hamiltonian, and the second is a counterpart of the Kitaev superconducting chain~\cite{kitaev01}. Finally, the Supplemental Material describes the mapping of $p=6$ parafermions into $p=3$ parafermions and fermions ($p=2$), and of $p=8$ parafermions into three species of fermions.

\bibitem{alicea12}
J. Alicea,
New directions in the pursuit of Majorana fermions in solid state systems,
Rep.\ Prog.\ Phys.\ {\bf 75}, 076501 (2012).

\bibitem{leijnse12}
M. Leijnse and K. Flensberg,
Introduction to topological superconductivity and Majorana fermions,
Semicond.\ Sci.\ Technol.\ {\bf 27}, 124003 (2012).

\bibitem{beenakker15}
C. W. J. Beenakker, 
Random-matrix theory of Majorana fermions and topological superconductors,
Rev. Mod. Phys. {\bf 87}, 1037 (2015).

\bibitem{guo16}
H.-M. Guo,
A brief review on one-dimensional topological insulators and superconductors,
Sci.\ China Phys.\ Mech.\ {\bf 59}, 637401 (2016).

\bibitem{sato17}
M. Sato and Y. Ando,
Topological superconductors: a review,
Rep.\ Prog.\ Phys.\ {\bf 80}, 076501 (2017).

\bibitem{teixeira21}
R. L. R. C. Teixeira and L. G. G. V. Dias da Silva,
Quantum dots as parafermion detectors,
Phys.\ Rev.\ Research {\bf 3}, 033014 (2021).

\bibitem{kitaev01}
A. Y. Kitaev,
Unpaired Majorana fermions in quantum wires,
Phys.-Usp.\ {\bf 44}, 131 (2001).

\bibitem{li15}
W. Li, S. Yang, H.-H. Tu, and M. Cheng,
Criticality in translation-invariant parafermion chains,
Phys.\ Rev.\ B {\bf 91}, 115133 (2015).

\bibitem{statphys}
M. P. Kennett,
{\it Essential Statistical Physics}
(Cambridge University Press, Cambridge, 2021).

\bibitem{dutt94}
Thermodynamics of a free $q$-fermion gas,
R. Dutt, A. Gangopadhyaya, A. Khare, and U. P. Sukhatme,
Int.\ J.\ Mod.\ Phys.\ A {\bf 9}, 2687 (1994).

\bibitem{stoilova20}
N. I. Stoilova and J. Van der Jeugt,
Partition functions and thermodynamic properties of paraboson and parafermion systems,
Phys.\ Lett.\ A {\bf 384}, 126421 (2020).

\bibitem{datanote}
\href{https://doi.org/10.17635/lancaster/researchdata/362}{https://doi.org/10.17635/lancaster/researchdata/362}

\end{thebibliography}

\begin{thebibliography}{99}

\bibitem{asboth16s}
J. K. Asb\'oth, L. Oroszl\'any, and A. P\'alyi,
{\it A Short Course on Topological Insulators}
(Springer, Switzerland, 2016).

\bibitem{kitaev01s}
A. Y. Kitaev,
Unpaired Majorana fermions in quantum wires,
Phys.-Usp.\ {\bf 44}, 131 (2001).

\bibitem{alicea12s}
J. Alicea,
New directions in the pursuit of Majorana fermions in solid state systems,
Rep.\ Prog.\ Phys.\ {\bf 75}, 076501 (2012).

\bibitem{leijnse12s}
M. Leijnse and K. Flensberg,
Introduction to topological superconductivity and Majorana fermions,
Semicond.\ Sci.\ Technol.\ {\bf 27}, 124003 (2012).

\bibitem{beenakker15s}
C. W. J. Beenakker, 
Random-matrix theory of Majorana fermions and topological superconductors,
Rev. Mod. Phys. {\bf 87}, 1037 (2015).

\bibitem{guo16s}
H.-M. Guo,
A brief review on one-dimensional topological insulators and superconductors,
Sci.\ China Phys.\ Mech.\ {\bf 59}, 637401 (2016).

\bibitem{sato17s}
M. Sato and Y. Ando,
Topological superconductors: a review,
Rep.\ Prog.\ Phys.\ {\bf 80}, 076501 (2017).

\bibitem{baxter89as}
R. J. Baxter,
A simple solvable $Z_N$ Hamiltonian,
Phys. Lett. A {\bf 140}, 155 (1989).

\bibitem{baxter89bs}
R. J. Baxter,
Superintegrable chiral Potts model: Thermodynamic properties, an “inverse” model, and a simple associated Hamiltonian.
J. Stat. Phys. {\bf 57}, 1 (1989).

\bibitem{fendley14s}
P. Fendley,
Free parafermions,
J. Phys. A: Math. Theor. {\bf 47}, 075001 (2014).

\bibitem{alicea16s}
J. Alicea and P. Fendley,
Topological Phases with Parafermions: Theory and Blueprints,
Annu.\ Rev.\ Condens.\ Matter Phys.\ {\bf 7}, 119 (2016).

\bibitem{alcaraz17s}
F. C. Alcaraz, M. T. Batchelor, and Z.-Z. Liu,
Energy spectrum and critical exponents of the free parafermion $Z_N$ spin chain,
J.\ Phys.\ A: Math.\ Theor.\ {\bf 50}, 16LT03 (2017).

\bibitem{batchelor23s}
M. T. Batchelor, R. A. Henry, and X. Lu,
A brief history of free parafermions,
AAPPS Bulletin {\bf 33}, 29  (2023).

\bibitem{cobanera14s}
E. Cobanera and G. Ortiz,
Fock parafermions and self-dual representations of the braid group,
Phys.\ Rev.\ A {\bf 89}, 012328 (2014).

\bibitem{calzona18s}
A. Calzona, T. Meng, M. Sassetti, and T. L. Schmidt,
$\mathbb{Z}_4$ parafermions in one-dimensional fermionic lattices,
Phys.\ Rev.\ B {\bf 98}, 201110(R) (2018).

\bibitem{teixeira21s}
R. L. R. C. Teixeira and L. G. G. V. Dias da Silva,
Quantum dots as parafermion detectors,
Phys.\ Rev.\ Research {\bf 3}, 033014 (2021).

\end{thebibliography}
\end{document}